\documentclass[aps,showpacs,nofootinbib%
,twocolumn]{revtex4}

\usepackage{graphicx}
\usepackage[mathscr]{eucal}
\usepackage{color}
\usepackage{graphicx}
\usepackage{epsf}
\usepackage{bm}
\usepackage{epsfig}
\usepackage{amsmath}
\usepackage[figuresleft]{rotating}
\usepackage{young}
\usepackage{amsfonts}

\usepackage{booktabs}
\usepackage{bm}

 \oddsidemargin  = - 7 mm 
 \evensidemargin = - 7 mm
\newcommand{\Bb}{{\bar{B}}}

\newcommand{\Bsb}{{\bar{B}_s}}

\newcommand{\BSb}{{\bar{B}^*}}

\newcommand{\BsSb}{{\bar{B}_s^*}}

\newcommand{\Lambdab}{{\Lambda_b}}

\newcommand{\HQSS}{{\rm HQSS}}
\newcommand{\SU}{{\rm SU}}

\newcommand{\MeV}{{\,\rm MeV}}

\newcommand{\TeV}{{\,\rm TeV}}

\newcommand{\ignore}[1]{} %VERY USEFUL. NEVER USE WITH verbatim

\input{epsf.tex}
\begin{document}

\title{Odd parity bottom-flavored baryon resonances}

\author{ C.~Garc{\'\i}a-Recio$^1$, J.~Nieves$^2$, O.~Romanets$^3$, 
  L.~L.~Salcedo$^1$, L.~Tolos$^{4,5}$}
\affiliation{
$^1$Departamento~de~F{\'\i}sica~At\'omica, Molecular~y~Nuclear, and Instituto
  Carlos I de F{\'\i}sica Te\'orica y Computacional, Universidad~de~Granada,
  E-18071~Granada, Spain\\
$^2$Instituto~de~F{\'\i}sica~Corpuscular~(centro~mixto~CSIC-UV),
  Institutos~de~Investigaci\'on~de~Paterna, Aptdo.~22085,~46071,~Valencia,
  Spain\\
$^3$KVI,
  University~of~Groningen, Zernikelaan~25,~9747AA~Groningen, The~Netherlands
  \\
$^4$Institut~de~Ci\`encies~de~l'Espai~(IEEC/CSIC),
  Campus~Universitat~Aut\`onoma~de~Barcelona, Facultat~de~Ci\`encies,
  Torre~C5,~E-08193~Bellaterra,~Spain \\
$^5$Frankfurt Institute for Advanced Studies, \\
Johann Wolfgang Goethe University, Ruth-Moufang-Str.~1,
60438 Frankfurt am Main
}

\date{\today}

\pacs{14.20.Lq, 11.10.St, 12.38.Lg, 14.40.Lb} 

\begin{abstract}
The LHCb Collaboration has recently observed two narrow baryon resonances with
beauty. Their masses and decay modes look consistent with the quark model
orbitally excited states $\Lambda_b(5912)$ and $\Lambda^*_b(5920)$, with
quantum numbers $J^P=1/2^-$ and $3/2^-$, respectively. We predict the
existence of these states within a unitarized meson-baryon coupled-channel
dynamical model, which implements heavy-quark spin symmetry. Masses, quantum
numbers and couplings of these resonances to the different meson-baryon
channels are obtained. We find that the resonances $\Lambda^0_b(5912)$ and
$\Lambda^0_b(5920)$ are heavy-quark spin symmetry partners, which naturally
explains their approximate mass degeneracy.  Corresponding bottom-strange
baryon resonances are predicted at $\Xi_b(6035.4)$ ($J^P=\frac{1}{2}^-$) and
$\Xi_b(6043.3)$ ($J^P=\frac{3}{2}^-$). 
The two $\Lambda_b$ and two $\Xi_b$
resonances complete a multiplet of the combined symmetry SU(3)-flavor times
heavy-quark spin.
\end{abstract}

\maketitle
%\tableofcontents

\section{Introduction}
Using $pp$ collision data at $7\TeV$ center of mass energy, the LHCb
Collaboration~\cite{Aaij:2012da} has reported the existence of two narrow
states observed in the $\Lambda^0_b\pi^+\pi^-$ spectrum, with masses $5911.95
\pm 0.12\,(stat)\,\pm 0.03\,(syst)\,\pm 0.66\,(\Lambda^0_b \, {\rm mass})
\MeV$, and $5919.76\pm 0.07\,(stat)\,\pm 0.02\,(syst)\, \pm
0.66\,(\Lambda^0_b\, {\rm mass})\MeV$. These states are interpreted as the
orbitally-excited $\Lambda^0_b(5912)$ and $\Lambda^0_b(5920)$ bottom baryon
resonances, with spin--parity $J^P=1/2^-$ and $J^P=3/2^-$, respectively.  The
limits on the natural widths of these states are $ \Gamma_{\Lambda^0_b(5912)}
\leq 0.82\MeV$ and $ \Gamma_{\Lambda^0_b(5920)} \leq 0.72\MeV$ at the 95\%
confidence level ~\cite{Aaij:2012da}.

There exists an old prediction by Capstick and Isgur for the masses of these
two $\Lambda_b$ resonances which is in very good agreement with the results
reported by the LHCb collaboration. Indeed, their relativistic quark model
predicts $5912 \MeV$ and $5920 \MeV$ for the masses of the lightest
orbitally-excited states~\cite{Capstick:1986bm}. However, the same model
yields a mass of the ground state $\Lambda^0_b$ ($J^P=1/2^+$) which is about
$35 \MeV$ smaller than the measured value~\cite{Beringer:1900zz}. More
recently, Garcilazo et al.~\cite{Garcilazo:2007eh} have also presented results
from a constituent quark model scheme.  They adjusted the mass of the
$\Lambda^0_b$ ground state and predicted the masses of the $J^P=1/2^-$ and
$3/2^-$ orbitally excited $\Lambda_b$ states, which turned out to be around
$30$ $\MeV$ lower than the LHCb experimental values. Note however that the
masses predicted in ~\cite{Garcilazo:2007eh} are in turn 20-$30\MeV$ higher
than those obtained in other schemes based also on the relativistic quark
model \cite{Ebert:2007nw}, or on the color hyperfine interaction
\cite{Karliner:2008sv} or on the heavy quark effective theory
\cite{Roberts:2007ni}. More recently, in \cite{Ortega:2012cx} heavy
  baryonic resonances $\Lambda_b$ ($\Lambda_c$) with $J^P=3/2^-$ are studied
  in a constituent quark model as a molecular state composed by nucleons and
  $\bar{B^*}$ ($D^*$) mesons.

In this work we adopt a different approach and describe these odd parity
excited states as dynamically generated resonances obtained within a
unitarized meson-baryon coupled-channel scheme. It is known that some baryon
states can be constructed as a $qqq$ state in a quark model, and
simultaneously as a dynamically generated resonance in a meson-baryon
coupled-channel description (that is a $qqq -q \bar q $ molecular state)
\cite{Klempt:2009pi}. Though, some of their properties might differ. It is
thus interesting to consider both points of view in order to get in the future
a joint or integral description of hadronic resonances in terms of quarks and
hadrons degrees of freedom.

The unitarization in coupled-channels has proven to be very successful in
describing some of the existing experimental data. Such studies include
approaches based on the chiral perturbation theory amplitudes for scattering
of $0^-$ octet Goldstone bosons off baryons of the $1/2^+$ nucleon octet in
the charmless sector
\cite{Kaiser:1995eg,Kaiser:1995cy,Oset:1997it,Krippa:1998us,Nacher:1999vg,Meissner:1999vr,Oller:2000fj,
  Jido:2003cb,Nieves:2001wt,Inoue:2001ip,Lutz:2001yb,Garcia-Recio:2002td,Oset:2001cn,Ramos:2002xh,Tolos:2000fj,GarciaRecio:2003ks,Oller:2005ig,Borasoy:2005ie,Borasoy:2006sr,Hyodo:2008xr}.
Unitarized coupled-channel methods have been further extended to the
baryon-meson sector with charm degrees of freedom under several variants: The
works
in~\cite{Tolos:2004yg,Lutz:2003jw,Lutz:2005ip,Hofmann:2005sw,Hofmann:2006qx,Lutz:2005vx,Mizutani:2006vq,Tolos:2007vh,JimenezTejero:2009vq}
use a bare baryon-meson interaction saturated with the $t$-channel exchange of
vector mesons between pseudoscalar mesons and baryons.  The works in
\cite{Haidenbauer:2007jq,Haidenbauer:2008ff,Haidenbauer:2010ch} are based on
the J\"ulich meson-exchange model. Those in
\cite{Wu:2010jy,Wu:2010vk,Oset:2012ap} apply the hidden gauge formalism, and
the same approach has been extended to the bottom sector in \cite{Wu:2010rv}.
Finally, an extended Weinberg-Tomozawa (WT) interaction is used in
\cite{GarciaRecio:2008dp,Gamermann:2010zz,Romanets:2012hm}.

Of special importance are the symmetries that are implemented in the quark or
hadronic models. Typically, while hadronic models pay an special attention to
chiral symmetry, quark models usually implement heavy-quark spin symmetry
(HQSS).  HQSS is a direct consequence of QCD
~\cite{Isgur:1989vq,Neubert:1993mb,MW00}. It states that the interaction
dependent on the spin state of the heavy quark is of $O(\Lambda_{{\rm
    QCD}}/m_Q)$, and so suppressed in the infinite quark mass limit. For
instance, the vector and pseudoscalar mesons with a bottom quark, which only
differ in how the spins of light and heavy quarks are coupled, form a doublet
of HQSS and would be degenerate in the infinite mass limit. So, HQSS requires
the pseudoscalar $\Bb~(\Bsb)$ meson and the $\BSb~(\BsSb)$ meson, its vector
partner, to be treated on an equal footing.  On the other hand, chiral
symmetry fixes the lowest order interaction between Goldstone bosons and other
hadrons in a model independent way; this is the WT interaction.  Thus, it is
appealing to have a predictive model for four flavors including all basic
hadrons (pseudoscalar and vector mesons, and $\frac{1}{2}^+$ and
$\frac{3}{2}^+$ baryons) which reduces to the WT interaction in the sector
where Goldstone bosons are involved and which incorporates HQSS in the sector
where bottom quarks participate.

In this letter we use hadronic degrees of freedom in a unitarized meson-baryon
coupled-channel calculation. We rely on a tree-level contact interaction that
embodies the approximate pattern of both chiral symmetry, when Goldstone
bosons are involved, and HQSS when heavy hadrons are present. Moreover, it
enjoys spin-flavor symmetry in the light ($u,~d,~s$) flavor sector.  The
scheme has been successfully used for describing odd parity $s$-wave light
flavor~\cite{Gamermann:2011mq} and
charm~\cite{GarciaRecio:2008dp,Gamermann:2010zz,Romanets:2012hm} baryon
resonances.  Indeed, the model naturally explains the overall features
(masses, widths and main couplings) of the corresponding resonances
($\Lambda_c^+(2595)$ and $\Lambda_c^+(2625)$) that appear in the charm sector
($C=1$)~\cite{GarciaRecio:2008dp,Romanets:2012hm}. Further predictions of this
model for the $C=-1$ sector can be found in
\cite{Gamermann:2010zz,GarciaRecio:2011xt}.

HQSS is not explicitly accounted for in other unitarized coupled-channel
models
\cite{Tolos:2004yg,Lutz:2003jw,Lutz:2005ip,Hofmann:2005sw,Hofmann:2006qx,%
  Lutz:2005vx,Mizutani:2006vq,Tolos:2007vh,JimenezTejero:2009vq,%
  Haidenbauer:2007jq,Haidenbauer:2008ff,Haidenbauer:2010ch,%
  Wu:2010jy,Wu:2010vk,Oset:2012ap,Wu:2010rv}, as they typically give an
asymmetric treatment to heavy mesons that are HQSS partners. Nevertheless, a
detailed analysis of the hidden gauge model as applied in
\cite{Wu:2010jy,Wu:2010vk,Wu:2010rv} shows no actual violation in the heavy
quark limit \cite{Eulogio}.\footnote{The would-be offending amplitudes are
  either not actually included due to the large gap between channels (e.g.,
  $\Lambda_b B \to N \pi$) or they would be suppressed in the heavy quark
  limit due to the presence of a heavy vector-meson exchange propagator (e.g.,
  $\Lambda_c \bar{D}\to N \eta_c$).}

The paper is organized as follows. In Sec.~\ref{model} we present the features
of our coupled-channel unitarized approach while in Sec.~\ref{results} we show
the results not only for ($\Lambda_b,\Lambda_b^*$) but also for
($\Xi_b,\Xi_b^*$) states, which belong to the same SU(3)$\times$HQSS
multiplets.  Finally, we present our conclusions in Sec.~\ref{conclusions}.

\section{Coupled-channels and unitarization}
\label{model}
We follow here the approach already applied in
Refs.~\cite{GarciaRecio:2008dp,Romanets:2012hm,Gamermann:2010zz} for charm
quarks. We will consider a system with baryon number one, one bottom quark
($B=-1$) and strangeness, isospin and spin-parity given by:
$(S,I,J^P)=(0,0,1/2^-)$ denoted as $\Lambda_b$, $(0,0,3/2^-)$ as
$\Lambda_b^*$, $(-1,1/2,1/2^-)$ as $\Xi_b$ and $(-1,1/2,3/2^-)$ as $\Xi_b^*$.

All meson-baryon pairs with the same $SIJ$ quantum numbers span
the coupled-channel space. We apply $s$-wave amplitudes. This seems appropriate
in our case since the observed resonances are close to threshold. The
 tree-level contact amplitudes between two channels $ij$ for
each $SIJ$ sector are given by:
\begin{equation}
V_{ij}^{SIJ} =
D_{ij}^{SIJ}\,\frac{2\sqrt{s}-M_i-M_j}{4f_if_j} 
\sqrt{\frac{E_i+M_i}{2M_i}}\sqrt{\frac{E_j+M_j}{2M_j}}, 
\label{eq:pot}
\end{equation}
where $\sqrt{s}$ is the center of mass (C.M.) energy of the system; $E_i$ and
$M_i$ are, respectively, the C.M. energy and mass of the baryon in the channel
$i$; and $f_i$ is the decay constant of the meson in the $i$-channel.  The
masses of baryons with bottom content used in this work are compiled in
Table~\ref{tab:baryonmass}, while those of the bottom mesons and their decay
constants are given in Table~\ref{tab:mesonmass}.  The rest of hadron masses
and meson decay constants not shown in the above tables have been taken from
Ref.~\cite{Romanets:2012hm}.

Finally the coefficients $D_{ij}^{SIJ}$ come from the underlying spin-flavor
extended WT structure of the couplings in our model
\cite{GarciaRecio:2005hy,GarciaRecio:2008dp,Romanets:2012hm}. Tables for the
coefficients can be found in the Appendices B of
Refs.~\cite{GarciaRecio:2008dp,Romanets:2012hm}. The coefficients to be used
for the $B=-1$ sector (one bottom quark interacting with light quarks) are
identical to those for $C=1$ (one charm quark interacting with light quarks)
with obvious renaming of the heavy hadrons. The universality of the
interactions of heavy quarks, regardless of their concrete (large) mass,
flavor and spin state, follows from QCD
\cite{Isgur:1989vq,Neubert:1993mb,MW00} and it is automatically implemented in
our model. Let us note that such emerging heavy spin-flavor symmetry, which
becomes exact in the infinitely heavy quark limit, is different from the
approximate SU(6) or light spin-flavor symmetry, also implemented in our
model.

\newpage
We remark that the various exact symmetries referred to above (chiral,
spin-flavor and HQSS) apply only to the coefficients $D_{ij}^{SIJ}$, while
physical masses and decay meson constants are used throughout when solving the
coupled-channel equations. The symmetry content of our model can be exposed by
artificially changing these hadron properties (masses and decay constants) to
enforce spin-flavor, flavor and/or heavy quark spin symmetries. If, starting
from the physical values, $\SU(3)\times\HQSS$ is adiabatically enforced, the
resonances so obtained organize themselves into exact $\SU(3)\times \HQSS$
multiplets. In this way this approximate symmetry can be used to label the
physical states. The largest exact symmetry present in the coefficients
$D_{ij}^{SIJ}$ is $\SU(6)\times \HQSS$, so the physical states can also be
classified under approximate multiplets of this symmetry.\footnote{ The
  requirement of $\SU(6)\times \HQSS$ still allows many possible interactions
  so spin-flavor SU(8) is used to reduce the number of parameters, but this
  symmetry is explictly broken, even at the level of coefficients, in order to
  have exact HQSS \cite{Romanets:2012hm}.} We apply such group label
assignments to our results.

We use the matrix $V^{SIJ}$ as kernel to solve the Bethe-Salpeter equation,
which provides the $T$-matrix as
\begin{equation}
T^{SIJ}=(1-V^{SIJ}G^{SIJ})^{-1}V^{SIJ}\label{eq:bse}.
\end{equation}
Here $G^{SIJ}$ is a diagonal matrix containing the two particle propagator
for each channel. Explicitly
\begin{equation}
G^{SIJ}_{ii} = 2M_i
\left(
\bar{J}_0(\sqrt{s};M_i,m_i) - \bar{J}_0(\mu^{SI};M_i,m_i)
\right)
, 
\label{eq:subs}
\end{equation}
where $m_i$ is the mass of the meson in the channel $i$. The loop function
$\bar{J}_0$ can be found in the Appendix of Ref.~\cite{Nieves:2001wt} for the
different relevant Riemann sheets. The two particle propagator diverges
logarithmically. The loop is renormalized by a subtraction constant such that
\begin{equation}
G_{ii}^{SIJ}=0 \quad\text{at~~} \sqrt{s}=\mu^{SI}. 
\label{eq:musi}
\end{equation}
To fix the subtraction point $\mu^{SI}$, we apply the following prescription:
$\mu^{SI}$ equals $\sqrt{m_{\rm{th}}^2+M_{\rm{th}}^2}$, where $m_{\rm{th}}$
and $M_{\rm{th}}$ are, respectively, the masses of the meson and the baryon of
the channel with the lowest threshold (minimal value of
$m_{\rm{th}}+M_{\rm{th}}$) among all the channels with the given values of $S$
and $I$ and any value of $J$.  Therefore the subtraction point takes a common
value for all sectors $SIJ$ with equal $SI$.  This renormalization scheme (RS)
was first proposed in \cite{Hofmann:2005sw,Hofmann:2006qx}.  Successful
results from this RS, but involving only the mesons and baryons of the pion
and nucleon octets were already obtained in \cite{GarciaRecio:2003ks}. This
specific RS is just a prescription which previously has produced good results,
but it can be refined when phenomenological information is
available.\footnote{See \cite{Hyodo:2008xr} for a discussion on the
  subtraction point and its natural vs. phenomenological values.} We will
study this possibility in the next section.

The dynamically-generated baryon resonances can be obtained as poles of the
scattering amplitudes in each of the $SIJ$ sectors.  We look at both the first
and second Riemann sheets of the variable $\sqrt{s}$. The poles of the
scattering amplitude on the first Riemann sheet that appear on the real axis
below threshold are interpreted as bound states. The poles that are found on
the second Riemann sheet below the real axis and above threshold are
identified with resonances.\footnote{Often we refer to all poles generically
  as resonances, regardless of their concrete nature, since usually they can
  decay through other channels not included in the model space.}  The mass and
the width of the bound state/resonance can be found from the position of the
pole on the complex energy plane. Close to the pole, the $T$-matrix behaves as
\begin{equation}
T^{SIJ}_{ij} (s) \approx \frac{g_i e^{i\phi_i}\,g_je^{i\phi_j}}{\sqrt{s}-\sqrt{s_R}}
\,.  
\label{Tfit} 
\end{equation}
$\sqrt{s_R}=M_R - \rm{i}\, \Gamma_R/2$ provides the mass ($M_R$) and the width
($\Gamma_R$) of the resonance, and $g_j e^{i\phi_j}$ (modulus and phase) is
the coupling of the resonance to the channel $j$.

\begin{table}%[h!]
\begin{center}
\begin{tabular}{| c l c r c c |}
\hline
 Baryon & ~$M\,[\MeV]$~& $\Gamma \, [\MeV]$
& ~$\SU(6)$ & $\SU(3)_{2J+1}$ & HQSS
\\
\hline
 $\Lambda_b$ & $5619.37$~\cite{note1}
& 
& $\bm{21}$ & $\bm{3^*_2}$ & singlet\\
 $\Xi_b$ & $5789.55$~\cite{Beringer:1900zz} & 
& $\bm{21}$ & $\bm{3^*_2}$ & singlet
\\
  $\Sigma_b$ & $5813.4\phantom{0}$~\cite{Beringer:1900zz} & $7.3$
& $\bm{21}$ & $\bm{6_2}$ & doublet
\\
  $ \Sigma_b^*$ & $5833.55$~\cite{Beringer:1900zz} & $9.5$
& $\bm{21}$ & $\bm{6_4}$ & doublet
\\
  $\Xi^\prime_b$ & $5926\phantom{.00}$~\cite{Ref:Xipb}
 & 
& $\bm{21}$ & $\bm{6_2}$ & doublet
\\
  $\Xi_b^*$ & $5945\phantom{.00}$~\cite{Chatrchyan:2012ni} & 
& $\bm{21}$ & $\bm{6_4}$ & doublet 
\\
  $\Omega_b$ & $6050.3\phantom{0}$~\cite{LHCb-CONF-2011-060} &
& $\bm{21}$ & $\bm{6_2}$ & doublet
\\
  $\Omega_b^*$ & $6069\phantom{.00}$~\cite{AliKhan:1999yb} &
& $\bm{21}$ & $\bm{6_4}$ & doublet
\\
\hline
\end{tabular}
\end{center}
\caption{Baryon masses and widths used throughout this work.  The $\SU(6)$ and
  $\SU(3)_{2J+1}$ labels are also displayed. The last column indicates the
  HQSS multiplets. Members of a HQSS doublet are placed in consecutive rows.}
\label{tab:baryonmass}
\end{table}
\begin{table}%[h!]

\begin{center}
\begin{tabular}{|c l c  r c c|}
\hline
Meson  & $m\,[\MeV]$ & $f\,[\MeV]$
& ~$\SU(6)$ & $\SU(3)_{2J+1}$ & HQSS 
\\
\hline
$\Bb$  & $5279.335$~\cite{Beringer:1900zz} & $133.6$~\cite{Na:2012kp}
& $\bm{6^*}$ & $\bm{3^*_1}$ & doublet
\\
$\BSb$ & $5325.2\phantom{00}$~\cite{Beringer:1900zz} & $f_\Bb$
& $\bm{6^*}$ & $\bm{3^*_3}$ & doublet
\\
$\Bsb$ & $5366.3\phantom{00}$~\cite{Beringer:1900zz} & $159.1$~\cite{McNeile:2011ng}
& $\bm{6^*}$ & $\bm{3^*_1}$ & doublet
\\
$\BsSb$ & $5415.4\phantom{00}$~\cite{Beringer:1900zz} & $f_\Bsb$
& $\bm{6^*}$ & $\bm{3^*_3}$ & doublet
\\
\hline
\end{tabular}
\end{center}
\caption{Meson masses, $m$, and decay constants,
  $f$, used throughout this work.  
The $\SU(6)$ and
  $\SU(3)_{2J+1}$ labels are also displayed. The last column indicates
  the HQSS multiplets. Members of a HQSS doublet are placed in consecutive rows.}
\label{tab:mesonmass}
\end{table}

There is a technical aspect which should be addressed at this point. It
follows from QCD that, as one flavor of quarks becomes heavy, the spectrum of
hadrons with one such quark tends to a universal pattern, shifted by the heavy
quark mass. However, it is well known \cite{Nieves:2001wt} that the
renormalized loop function, $G$, grows logarithmically as any one of the
hadrons in the loop gets heavy. This implies that, in the infinitely heavy
quark limit, the interaction (and so the binding energy in attractive sectors)
would effectively increase at a logarithmic rate, rather than stabilizing. By
artificially increasing the bottomed hadron masses we have verified that such
spurious binding would indeed arise for sufficiently large masses,\footnote{
  Simultaneously, we find that the gaps between resonances decrease as
  $1/m_Q$.}  however, it is not clear how sizable the effect is in a realistic
scenario.  As will be seen below, the generic subtraction point defined after
Eq.~(\ref{eq:musi}) actually produces too little binding and we have to move
to a phenomenological subtraction point to pinpoint the experimentally
observed states. This would suggest that the problem is not yet a pressing one
at the bottom scale, at least for the sector we are considering and those
related to it by softly broken symmetries. It can be expected that whenever
the subtraction point is shifted to fine tune the overall position of a
multiplet of resonances, any spurious binding will produce at most a residual
distortion in the individual positions, without compromising the existence and
main couplings of the resonances under study.

\section{Results and Discussion}
\label{results}

\subsection{$\Lambda_b$ and $\Lambda_b^*$ states}

%%%%%%%%%%%%%%%%%%%%%%%% TABLE 1
\begin{table*}[ht]
\begin{center}
\begin{tabular}{| c | c| c | c | c | c | c | c | c | c |}
\hline 
 SU(6) & SU(3)$_{2J+1}$  & $\Delta M_R$ & $M_{R}$ & $\Gamma_{R}$   & Couplings  	 &  &
 Experimental & Decay \\
 irrep             & irrep    & MeV &  MeV  & &  to main
 channels  & $J$  & LHCb & mode  \\ 
\hline
$\bf{21}$ &  $\bf{3_2^*}$ & 178 & ~5797.6~ & 0 &   $g_{N \bar B}=4.9$,   $g_{N \bar B^*}=8.3$,  
       $g_{\Lambda \bar B_s^{0}}=2.1$,    $g_{\Lambda \bar B_s^{*}}=3.6$,   &
       $1/2$  & & $\Lambda_b\gamma$ \\
&  &  &  &   &   $g_{\Lambda_b \eta'}=1.0$,   $g_{\Sigma_b^* \rho}=0.6$    &
       &   & \\
\hline

 $\bf{15}$  &  $\bf{3_2^*}$ & 291  & 5910.1  & 0 &  $g_{\Sigma_b \pi}=1.8$,   $g_{N \bar B}=4.6$,
    $g_{N \bar B^*}=3.0$,    $g_{\Lambda_b \omega}=1.4$       & 1/2  &
    $\Lambda_b(5912)$ & $\Lambda_b\pi\pi$\\
\hline

 $\bf{15}$  &  $\bf{3_4^*}$  & 301 &  5921.5 &  0  &  $g_{\Sigma_b^* \pi}=1.8$,   $g_{N \bar B^*}=5.7$, 
    $g_{\Lambda_b \omega}=1.5$                      &  3/2  &
    $\Lambda^*_b(5920)$  & $\Lambda_b\pi\pi$\\

\hline

 $\bf{21}$  &   $\bf{3_2^*}$  & 390  &  6009.3  &  0 &     $ \bf{g_{\Sigma_b \pi} \sim 0.05}$,   $g_{\Lambda_b \eta}=2.0$,    $g_{N \bar B}=1.1$,  
    $g_{N \bar B^*}=1.7$,    & 1/2  &  & $\Sigma_b\pi$\\

&  &  &  &   & $g_{\Xi_b K}=0.8$,  $g_{\Lambda \bar B_s^{0}}=3.9$,  $g_{\Lambda \bar B_s^{*}}=6.0$,
   $g_{\Sigma_b^* \rho}=0.7$,    &  & & \\

&  &  &  &   &   $g_{\Xi_b^* K^*}=0.9$  &  & & \\

\hline

\end{tabular}
\end{center}
\caption{$\Lambda_b$ ($J^P=1/2^-$) and $\Lambda_b^*$ ($J^P=3/2^-$) resonances
  predicted in this work. The parameter $\alpha$ in Eq.~(\ref{eq:defalpha})
  has been set to 0.967. The SU(6) and $\SU(3)\times\SU(2)$ representations of
  the corresponding states are shown in the first two columns.  $M_R$,
  $\Gamma_R$ and $\Delta M_R$ stand for the mass, the width, and the mass
  difference with respect to the ground state $(J^P=1/2^+)$ $\Lambda_b$.  The
  next column displays the (absolute value of the) dominant couplings to the
  different meson-baryon channels, ordered by their threshold energies.  The
  couplings to channels open for decay are highlighted in bold font.  The
  seventh column shows the spin of the resonance.  Tentative identifications
  with experimental resonances of LHCb are also given in the following
  column. Finally, in the last column we show the decay channel with largest
  phase-space allowed by strong interactions (or electromagnetic ones when the
  strong decay is forbidden). The two states in the {\bf 15} of SU(6) form a
  HQSS doublet, the other two states are HQSS singlets. The three lightest
  states belong to the {\bf 168} of SU(8), the heaviest one belongs to the
  {\bf 120} (note that it is precisely in these two SU(8) irreps where the WT
  interaction is more attractive, and thus the lowest-lying states stem from
  them~\cite{Romanets:2012hm}).}
\label{tablambdab}
\end{table*}
%%%%%%%%%%%%%%%

In the $\Lambdab$ sector ($B=-1,~C=0,~S=0,~I=0,~J^P=1/2^-$),
 the following sixteen channels are
involved:
\noindent
\begin{center}
\begin{tabular}{llllllll}
$\Sigma_b \pi $   & $\Lambda_b \eta$   & $N \bar B$  & $N \bar B^*$    
& $\Xi_b K$    &  $\Lambda_b \omega$     &  $ \Xi'_b K$  & $ \Lambda \bar B_s^{0} $   \\
$ \Lambda \bar B_s^{*}$  &   $\Lambda_b \eta'$    &  $\Sigma_b \rho$    &  $\Sigma_b^* \rho$   & $\Lambda_b \phi$    
& $\Xi_b K^*$    & $\Xi'_b K^*$     & $\Xi_b^* K^*$  \\
\end{tabular}
\end{center}
%\smallskip
%
Likewise for the $\Lambda^*_b$ sector ($B=-1,~C=0$, $S=0$, $I=0$, $J^P=3/2^-$),
there are eleven channels:
\noindent
\begin{center}
\begin{tabular}{llllll}
$ \Sigma_b^* \pi $   & $ N \bar B^*  $  & $ \Lambda_b \omega  $   & $ \Xi_b^* K  $    
& $ \Lambda \bar B_s^*  $    & $ \Sigma_b \rho  $  \\
 $ \Sigma_b^* \rho  $  & $ \Lambda_b \phi  $   & $ \Xi_b K^* $ 
& $ \Xi'_b K^*  $   & $  \Xi_b^* K^*  $ &  \\
\end{tabular}
\end{center}
In both cases the channels are ordered by increasing mass thresholds.
                     
By solving the coupled-channel Bethe-Salpeter equation several states are
generated in each of the two sectors. The three lowest lying $\Lambda_b$
resonances have masses of $5880$ and $5949\MeV$ ($J^P=1/2^-$) and $5963\MeV$
($J^P=3/2^-$).  As one can expect, the situation in the $J=1/2^-$ channel
keeps a close parallelism with that of the $\Lambda_c(2595)$ resonance in the
charm sector~\cite{GarciaRecio:2008dp, Romanets:2012hm}. For both heavy
flavors the structure obtained mimics the well-known two-pole pattern of the
$\Lambda(1405)$
\cite{Jido:2003cb,GarciaRecio:2003ks,Garcia-Recio:2002td}. Thus, we find that
the state at $5880$ strongly couples to the $N \bar B$ and $N \bar B^*$
channels, with a negligible $\Sigma_b \pi $ coupling, while the $5949\MeV$
state has a sizable coupling to this latter channel. On the other hand, the
$J^P=3/2^-$ state at $5963$ is generated mainly by the ($N\bar{B}^*$,
$\Sigma_b^* \pi$) coupled-channel dynamics. This state is the bottom
counterpart of the $\Lambda(1520)$ and $\Lambda_c^*(2625)$ resonances.

These results are encouraging, but to achieve a better description of the
$\Lambda_b(5912)$ and $\Lambda_b(5920)$ states reported by the LHCb
Collaboration, we have slightly changed the value of the subtraction point
used in the RS defined by Eqs.~(\ref{eq:subs}) and (\ref{eq:musi})
\cite{GarciaRecio:2008dp}.  Thus, in this sector, we have set the meson-baryon
loop to be zero at the C.M. energy $\sqrt{s}=\mu$ given by
\begin{equation}
\mu^2 =\alpha~(M_{\Sigma_b}^2+m_\pi^2) \  . \label{eq:defalpha}
\end{equation}
For $\alpha=0.967$, we find two poles above the $\Lambda^0_b\pi\pi$ threshold,
with masses $5910.1\MeV$ ($J^P=1/2^-$) and $5921.5\MeV$ ($J^P=3/2^-$), which
admit a natural identification with the two experimental $\Lambda_b$
resonances observed in~\cite{Aaij:2012da}. The results for masses, widths and
couplings are presented in the Table~\ref{tablambdab}.  We have assigned
well-defined group labels to the resonances.  The multiplets of
$\SU(6)\times\HQSS$ and of $\SU(3)\times\HQSS$ to which the resonances belong
are identified by means of the procedure discussed at length in
Ref.~\cite{Romanets:2012hm}, namely, by adiabatically following the
trajectories of the poles generated as the various symmetries are restored or
broken.
The mass differences $\Delta M_R$ of the resonances with respect of ground
state $\Lambda_b$ are also shown in Table~\ref{tablambdab}.
\begin{equation}
 \Delta M_R=M_R-M_{\Lambda_b(g.s.)}
\end{equation}
For each resonance, the decay mode with largest phase-space allowed by strong
interactions (or electromagnetic ones when the strong decay is forbidden) is
shown in the last column.

We find that the states $\Lambda_b(5912)$ and $\Lambda^*_b(5920)$ are
heavy-quark spin symmetry partners. Indeed, these two states would be part of
a {\bf $3^*$} irreducible representation (irrep) of SU(3), embedded in a {\bf
  15} irrep of SU(6) (which in turn belongs to the irrep {\bf 168} of SU(8)
~\cite{Romanets:2012hm}). Thus, the light quark structure of these two states
is the same, and in particular their total spin, $s_l=1$. Hence, the coupling
of the $b$-quark spin ($j_b=1/2$) with the spin of the light degrees of
freedom yields $J=1/2$ and $J=3/2$. Then the two states, $\Lambda_b(5912)$ and
$\Lambda^*_b(5920)$, form an approximate degenerate doublet; they are
connected by a spin rotation of the $b$-quark.

Comparison of Table \ref{tablambdab} with the Table III of
Ref.~\cite{Romanets:2012hm} in the charm sector, shows that states with the
same group labels in both tables are the heavy flavor counterpart of each
other. In particular, the $\Lambda_b(5920)$ resonance is the bottom version of
$\Lambda_c(2625)$ one, while the $\Lambda_b(5912)$ would not be the
counterpart of the $\Lambda_c(2595)$ resonance, but it would be of the second
charmed state that appears around $2595\MeV$, and that gives rise to the two
pole structure~\cite{Romanets:2012hm} mentioned above. The same conclusion
follows from inspection of their couplings: the $\Lambda_c(2595)$ couples
weakly to $\Sigma_c \pi$ while the coupling to $\Sigma_b\pi$ is sizable for
the $\Lambda_b(5912)$ state.

The two states observed by the LHCb Collaboration are detected through their
decay to $\Lambda_b({\rm g.s.})\pi\pi$. 
The fit to the data of the experiment of Ref.~\cite{Aaij:2012da} yields 
\begin{equation}
N(pp\to\Lambda^*_b(5920)\to\Lambda_b\pi\pi)=16.4\pm4.7
\end{equation}\
events with mass $M_{\Lambda_b(5912)} = 5911.95 \pm 0.11\MeV$ and 
\begin{equation}
N(pp\to\Lambda_b (5912)\to\Lambda_b\pi\pi)=49.5 \pm 7.9
\end{equation}
events with mass $M_{\Lambda^*_b(5920)} = 5919.76\pm 0.07\MeV$.  The
experimental setup of LHCb and the strong decay mechanism of the resonances
observed, guarantees that the decay to $\Lambda_b\pi\pi$ always takes place
within the space and time intervals set for detection \cite{Anton}. Therefore
no bias is expected from the possible different decay rates of the two
resonances, and
\begin{equation}
\frac{N(pp\to\Lambda^*_b(5920))}{N(pp\to\Lambda_b(5912))}
=
\frac
{N(pp\to\Lambda^*_b(5920)\to\Lambda_b\pi\pi)}
{N(pp\to\Lambda_b (5912)\to\Lambda_b\pi\pi)}
.
\end{equation}
This translates into an experimental ratio of cross sections
\begin{equation}
\left.
\frac
{\sigma(pp\to\Lambda^*_b(5920))}
{\sigma(pp\to\Lambda_b (5912))}
\right|_{\rm exp}
=\frac{N(pp\to \Lambda^*_b(5920))}{N(pp\to \Lambda_b(5912))}
=3.0\pm 1.0
\label{eq:3pm1}
\end{equation}

From the theoretical side, due to the dominant strong interactions taking
place during creation and hadronization of the quark $b$, a natural assumption
is that the $b$-quark spin ends up in a random state. In that case, and
assuming that $\Lambda_b(5912)$ and $\Lambda^*_b(5920)$ form a HQSS doublet,
the ratio of production of these states should be the quotient of
multiplicities, that is:
\begin{equation}
\frac{\sigma(pp\to\Lambda^*_b(5920))}{\sigma(pp\to\Lambda_b(5912))}
\approx 
\frac{2J_{\Lambda^*_b}+1}{2J_{\Lambda_b}+1}=2 
.
\label{eq:pre}
\end{equation}
Although not fully satisfactory, this ratio is not inconsistent with the
observed ratio, in Eq.~(\ref{eq:3pm1}), and it gives support to our conclusion
that the two observed states form a HQSS doublet.

From the couplings shown in Table \ref{tablambdab}, the dominant decay
mechanism of $\Lambda_b(5912)$ is expected to be of the form
$\Lambda_b(5912)\to \Sigma_b \pi$ with subsequent decay of the off-shell heavy
baryon, $\Sigma_b \to \Lambda_b \pi$. Its heavy quark partner follows a
similar pattern with $\Sigma_b^*$ and $\Lambda_b^*$. The approximate HQSS
requires the two resonances to have a similar width. In order to estimate this
width, we consider the following effective Lagrangian
\begin{equation}
{\mathcal L}(x) =
 \frac{g_{\Sigma_b\pi}}{\sqrt{3}} \vec{\Sigma}_b^\dagger\vec{\pi}\Lambda^{\rm res}_b
 +
g\, \vec{\Sigma}_b^\dagger\sigma_i \partial_i\vec{\pi} \, \Lambda_b
+
{\rm h.c.}
\end{equation}
The averaged experimental decay width of the $\Sigma_b$, $7.3\MeV$, allows to
extract the value $g \approx 51$. The value of $g_{\Sigma_b\pi}=1.8$ taken
from our calculation, Table \ref{tablambdab}, gives a small width for
$\Lambda_b(5912)$ around $8\,{\rm keV}$. A similar calculation for
$\Lambda^*_b(5920)$ yields a width around $12\,{\rm keV}$.\footnote{These
  numbers are just estimates. Being close to threshold any refinement in the
  treatment will induce relatively large changes in the values quoted.} The
smallness of the widths are due to the reduced phase space available since the
resonances are fairly close to the threshold. This is consistent with the
experimental bounds quoted in \cite{Aaij:2012da}.
 
Different quark models
\cite{Capstick:1986bm,Garcilazo:2007eh,Ebert:2007nw,Karliner:2008sv,Roberts:2007ni}
have also conjectured the existence of one or more excited $\Lambda_b(1/2^-)$
and $\Lambda_b(3/2^-)$ states. While the predicted masses for
\cite{Garcilazo:2007eh,Ebert:2007nw,Karliner:2008sv,Roberts:2007ni} differ few
tenths of MeV from the LHCb experimental ones (see Table VIII of
Ref.~\cite{Roberts:2007ni} for a summary of some of the results), the early
work of Capstick and Isgur \cite{Capstick:1986bm} generated the first two
excited $\Lambda_b(1/2^-)$ and $\Lambda_b(3/2^-)$ states with masses that are
in very good agreement with the ones observed by the LHCb collaboration.
However, the ground state $\Lambda_b(1/2^+)$ mass in this scheme is below the
experimental one. Our model reproduces the experimental $\Lambda_b(5912)$ and
$\Lambda_b(5920)$ but with an alternative explanation of their nature as
molecular states, which moreover are HQSS partners.

%%%%%%%%%%%%%%%%%%%%%%%% TABLE Xi_b new (alpha=1)
\begin{table*}[ht]
\begin{center}
\begin{tabular}{| c |c| c | c | c | c | c | c | c|}
\hline 
 SU(6)  & SU(3)$_{2J+1}$  & $\Delta M_{R}$ & $M_{R}$   & $\Gamma_{R}$  & Couplings  	 &  &
 Main decay \\
 irrep             & irrep & MeV   & MeV   & MeV &  to main
 channels  & $J$ & mode \\ 
\hline

$\bf{21}$ &  $\bf{3_2^*}$ & 255   & ~5874.~  & 0.  &   $g_{\Lambda \bar B}=1.3$,  $g_{\Sigma \bar B}=4.4$,
  $g_{\Lambda \bar B^*}=2.3$,  $g_{\Sigma \bar B^*}=7.3$,   & $1/2$  &
                                                                                                                         ${\Xi_b\gamma}$ \\
 &  &  &  &   &   $g_{\Xi \bar B_s}=2.6$,   $g_{\Xi_b \eta'}=1.0$,  $g_{\Xi \bar B_s^*}=4.5$     &  & \\
\hline

 $\bf{15}$  &  $\bf{3_2^*}$ & 416   &  6035.4  & 0.  & $\bf{g_{\Xi_b \pi }\sim 0.05}$,  $g_{\Sigma_b \bar K}=2.3$,
  $g_{\Lambda \bar B}=1.$,
   $g_{\Sigma \bar B}=4.5$,  & 1/2                  & ${\Xi_b\pi}$ \\
 &  &  &  &   &  $g_{\Sigma \bar B^*}=2.8$,     $g_{\Xi_b \omega}=1.2$,  
$g_{\Sigma^* \bar B^*}=2.3$   & & \\
\hline

 $\bf{15}$  &  $\bf{3_4^*}$ & 424 & 6043.3  & 0.  &  $g_{\Sigma_b^* \bar K}=2.3$,
        $g_{\Lambda \bar B^*}=1.1$,  $g_{\Sigma \bar B^*}=5.5$,
        $g_{\Sigma^* \bar B}=1.4$,        &  3/2
        & $\Xi_b\pi$ \\
 &  &  &  &   &    $g_{\Xi_b \omega}=1.2$,         $g_{\Sigma^* \bar B^*}=1.7$    & & \\
\hline

 $\bf{21}$  &   $\bf{3_2^*}$  & 453    &  6072.8   & 0.3  &   $\bf{g_{\Xi_b \pi}=0.1}$, 
  $\bf{g_{\Xi'_b \pi}=0.1}$, 
   $g_{\Xi_b \eta}=2.4$,   $g_{\Lambda \bar B}=1.4$, 
      & 1/2  & ${\Xi_b\pi}$, ${\Xi'_b\pi}$ \\
      &  &  &  &   &     $g_{\Lambda \bar B^*}=2.3$,      $g_{\Sigma \bar B}=1.1$,  
$g_{\Sigma \bar B^*}=1.6$,
  $g_{\Xi \bar B_s}=2.9$,
   & &\\
 &  &  &  &   &     $g_{\Xi \bar B_s^*}=4.5$ 
  & &\\
\hline

\end{tabular}
\end{center}

\caption{SU(3) partners of the states in Table \ref{tablambdab}. Predictions
  for $\Xi_b$ ($J^P=1/2^-$) and $\Xi_b^*$ ($J^P=3/2^-$) resonances (with
  $\alpha=1 $ in Eq.~(\ref{eq:defalpha})). No experimental excited $\Xi_b$ or
  $\Xi^*_b$ resonances have been detected yet. The two states in the {\bf 15}
  of SU(6) form a HQSS doublet.  }
\label{tabxib}
\end{table*}
%%%%%%%%%%%%%%%

\subsection{${\bm \Xi_b}$ and ${\bm \Xi^*_b}$  states} 

Next, we analyze the $(B = -1,~C=0,~ S=-1~,I=1/2)$ sector, for both $J=1/2$
and $J=3/2$ spins ($\Xi_b$ and $\Xi^*_b$ states, respectively). Our model
predicts the existence of nine states (6 $\Xi_b$ and 3 $\Xi^*_b$) stemming
from the strongly attractive {\bf 120} and {\bf 168} SU(8) irreducible
representations (see Ref.~\cite{Romanets:2012hm} for the analogous charm
sector). However, only three $\Xi_b$ and one $\Xi^*_b$ belong to the same
SU(3)$\times$HQSS multiplets of the $\Lambda_b$ and $\Lambda^*_b$ states
reported in Table~\ref{tablambdab}. In this exploratory study, we restrict our
discussion only to these states.

In the $\Xi_b$ sector, the following thirty one channels are involved:
%\smallskip
\noindent
\begin{center}
\begin{tabular}{llllllll}
$\Xi_b \pi$   & $\Xi'_b \pi$   & $\Lambda_b \bar K$   & $\Sigma_b \bar K$ 
  & $\Xi_b \eta$   & $\Lambda \bar B$    & $\Lambda \bar B^*$   & $\Sigma \bar B$  \\
 $\Xi'_b \eta$   & $\Lambda_b \bar K^*$   & $\Sigma \bar B^*$   & $\Omega_b K$  
& $\Xi_b \rho$  & $\Xi_b \omega$   & $\Xi \bar B_s$  & $\Xi'_b \rho$ \\
$\Sigma_b \bar K^*$     & $\Xi'_b \omega$   & $\Sigma^* \bar B^*$    & $\Xi_b^*\rho$      
  & $\Sigma_b^* \bar K^*$    & $\Xi_b^* \omega$     & $\Xi \bar B_s^*$    & $\Xi_b \eta'$    \\
 $\Xi_b \phi$   & $\Xi'_b \eta'$   & $\Omega_b K^*$   & $\Xi'_b \phi$
& $\Xi^* \bar B_s^*$   & $\Omega_b^* K^*$   & $\Xi_b^* \phi$  \\
\end{tabular}
\end{center}
\noindent
while in the $\Xi^*_b$ sector, the twenty six channels, ordered by increasing thresholds, are:
\smallskip
\noindent
\begin{center}
\begin{tabular}{llllllll}
$\Xi_b^* \pi$    & $\Sigma_b^* \bar K$    & $\Lambda \bar B^*$        & $\Xi_b^* \eta$    
  & $\Lambda_b \bar K^*$     & $\Sigma \bar B^*$        & $\Omega_b^* K$               & $\Xi_b \rho$   \\
     $\Xi_b \omega$       & $\Sigma^* \bar B$         & $\Xi'_b \rho$        & $\Sigma_b \bar K^*$  
 & $\Xi'_b \omega$    & $\Sigma^* \bar B^*$     & $\Xi_b^* \rho$     & $\Sigma_b^* \bar K^*$   \\
 $\Xi_b^* \omega$    & $\Xi \bar B_s^*$    & $\Xi_b \phi$    & $\Xi^* \bar B_s$
   & $\Xi_b^* \eta'$   & $\Omega_b K^*$    & $\Xi'_b \phi$      & $\Xi^* \bar B_s^*$  \\
 $\Omega_b^* K^* $    & $\Xi_b^* \phi$  \\
\end{tabular}
\end{center}

%%%%%%%%%%%%%%%%%%%%%%%% TABLE

For the subtraction point we use $\mu^2 = M_{\Xi_b}^2+m_\pi^2$, and thus we
assume our default value $\alpha=1$ in Eq.~(\ref{eq:defalpha}). There is no
particularly good reason to use the same value as in the $\Lambda_b$
case. Even SU(3) does not relate the two $\mu^{SI}$ points,
$M_{\Sigma_b}^2+m_{\pi}^2$ and $M_{\Xi_b}^2+m_\pi^2$, required to fix the RS
in each sector. If instead we take $\alpha=0.967$ as in the $\Lambda_b$
sector, $\Xi_b$ and $\Xi_b^*$ binding energies (masses) will be larger
(smaller) by about $60\mbox{-}80\MeV$. These $60\mbox{-}80\MeV$ should be
admitted as an intrinsic systematic uncertainty in our predictions in this
sector.

In this way, we find the $\Xi_b$ and $\Xi^*_b$ states that complete the
$\Lambda_b$ and $\Lambda^*_b$ SU(3)$\times$HQSS multiplets. The properties of
the dynamically generated $\Xi_b$ and $\Xi^*_b$ states are compiled in
Table~\ref{tabxib}. By studying the evolution of the poles from the
SU(6)$\times$HQSS symmetric point, we find that $\Lambda_b(5797.6)$ and
$\Xi_b(5874)$ belong to the same irreducible representation, and similarly the
$\Lambda_b(6009.3)$ and $\Xi_b(6072.8)$ states. Also, the pair $\Xi_b(6035.4)$
and $\Xi_b^*(6043.3)$, in the {\bf 15} irrep of SU(6), form the HQSS doublet
related by SU(3) to the doublet formed by the $\Lambda_b(5910.1)$ and
$\Lambda_b^*(5921.5)$ states.

The three $\Xi_b$ and one $\Xi^*_b$ states have also partners in the charm
sector. We find that states with the same group labels are the heavy flavor
counterpart of each other, as already noted for the $\Lambda_b$ and
$\Lambda^*_b$ sectors. By comparing Table~\ref{tabxib} with Table V of
Ref.~\cite{Romanets:2012hm}, we see that the HQSS partners in the charm sector
coming from the ${\bf 15}$ representation, $\Xi_c(2772.9)$ and
$\Xi^*_c(2819.7)$, are the bottom counterparts of the $\Xi_b(6035.4)$ and
$\Xi^*_b(6043.3)$ states. Moreover, the charmed $\Xi_c(2699.4)$ and
$\Xi_c(2775.4)$ resonances are analogous to the $\Xi_b(5874)$ and
$\Xi_b(6072.8)$ ones in the bottom sector, respectively. None of these
bottomed states have been seen experimentally yet. Schemes based on quark
models
\cite{Capstick:1986bm,Garcilazo:2007eh,Ebert:2007nw,Karliner:2008sv,Roberts:2007ni}
predict $\Xi_b(1/2^-)$ and $\Xi_b(3/2^-)$ states with similar masses to our
estimates, though there exist some differences between the various
predictions.  The experimental observation of the $\Xi_b$ and $\Xi_b^*$
excited states and their decays might, on the other hand, provide some
valuable information concerning the nature of these states, whether they can
be described as pure quark states or they have an important molecular
component.

\begin{figure}[t]
\centerline{ 
\epsfysize = 65 mm  \epsfbox{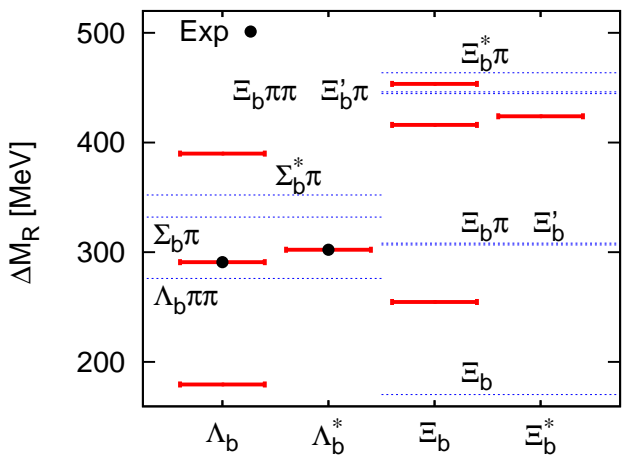}
      }
%\vspace{0.3cm}
\caption{Summary of the new predicted states. We also show the experimentally
  observed $\Lambda_b^0(5912)$ and $\Lambda_b^0(5920)$ states and some
  relevant hadronic thresholds. }
\label{fig:fig1}
\end{figure}

Fig.~\ref{fig:fig1} shows a summary of the masses of the predicted
$\Lambda_b(1/2^-)$, $\Lambda_b(3/2^-)$, $\Xi_b(1/2^-)$ and $\Xi_b(3/2^-)$
states with respect to the mass of the ground state $\Lambda_b$, together with
several thresholds for possible two- and three-body decay channels. The
experimental $\Lambda_b^0(5912)$ and $\Lambda_b^0(5920)$ of LHCb are given for
reference. Tables \ref{tablambdab} and \ref{tabxib} show that, except for
$\Xi_b(6072.8)$, our predicted states have a negligible width. This implies
that they do not strongly couple to two-body channels with lower mass, such as
$\Sigma_b \pi$ or $\Xi_b \pi$. Three-body channels are not included in our
calculation. These channels allow the possibility of strong decay for some of
the states. This is the case of the $\Lambda_b(3/2^-)$ and the two
$\Lambda_b(1/2^-)$ which lie above the threshold of $\Lambda_b \pi \pi$, but
it is not the case for the lightest $\Lambda_b$ and $\Xi_b$ states. They are
below all hadronic channels, and hence they are stable under strong
interactions. These states could be detected through electric dipole decay to
$\Lambda_b\gamma$ and $\Xi_b\gamma$.  Note that the strong decay of
$\Xi_b(3/2^-)$ to $\Xi_b\pi$ is forbidden in $s$-wave but allowed through
$d$-wave mechanisms not included in our model.

\begin{figure}[t]
\centerline{ 
\epsfysize = 55 mm  \epsfbox{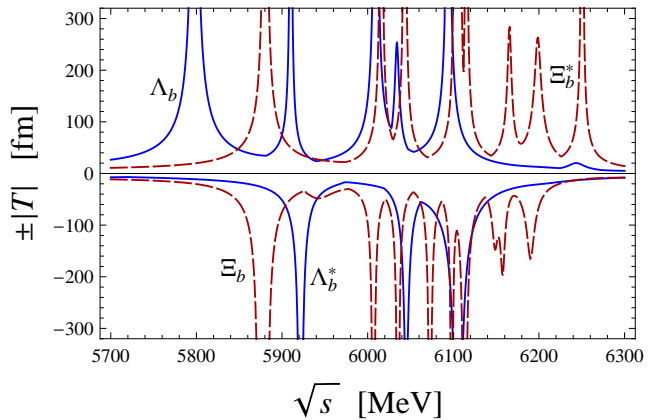}
      }
%\vspace{0.3cm}

\caption{ $\pm\max_j\sum_i|T_{ij}(\sqrt{s})|$ for the four $(I,J)$ sectors
  studied. We use "plus" sign for the sectors $\Lambda_b$ (blue, solid line)
  and $\Xi_b^*$ (red, dashed line) and "minus" sign for $\Lambda_b^*$ (blue,
  solid line) and $\Xi_b$ (red, dashed line). Exact SU(3) or HQSS symmetries
  would translate into exact mirror symmetries in the plot. }
\label{fig:fig2}
\end{figure}

In Fig.~\ref{fig:fig2}, we depict $(\pm |T|)-$matrix for the four $(SIJ)$
sectors studied in this work. Sectors related through SU(3) or by HQSS are
plotted with opposite sign to better appreciate the degree of fulfillment or
breaking of these symmetries. The extra poles stand for other states which
stem from other SU(8)/SU(6) irreps to those considered in this exploratory
study.

\section{Conclusions}
\label{conclusions}

We have analyzed odd parity baryons with one bottom quark by means of a
unitarized meson-baryon coupled-channel model which implements heavy-quark
spin symmetry. In particular, pseudoscalar and vector heavy mesons are treated
on an equal footing. We rely on a relatively simple tree-level contact
interaction already used in the charm sector
\cite{GarciaRecio:2008dp,Romanets:2012hm}: the matrix elements follow from
Clebsch-Gordan coefficients of the underlying spin-flavor symmetry with no
free-parameters. This interaction has the virtue of embodying the approximate
patterns of chiral symmetry, when Goldstone bosons are involved, and HQSS when
heavy hadrons are present.

A summary of our predictions is graphically shown in Fig.~\ref{fig:fig1}.  The
experimental states $\Lambda^0_b(5912)$ and $\Lambda^0_b(5920)$ reported by
the LHCb collaboration are obtained as dynamically generated meson-baryon
molecular states.  Within our scheme, these states are identified as HQSS
partners, which naturally explain their approximate mass degeneracy. Other
$\Lambda(1/2^-)$ states coming from the same attractive SU(6) $\times$ HQSS
representations are also analyzed and we find a close analogy to the charm and
strange sectors. In particular, the $\Lambda^0_b(5920)$ is the bottomed
counterpart of the $\Lambda^*(1520)$ and $\Lambda^*_c(2625)$
resonances. Moreover, the $\Lambda^0_b(5912)$ is part of a two-pole structure
similar as the one observed in the case of the $\Lambda(1405)$ and
$\Lambda_c(2595)$ resonances.

Mass and decay mode predictions are also obtained for some $\Xi_b(1/2^-)$ and
$\Xi_b(3/2^-)$ resonances, which belong to the same SU(3) multiplets as the
$\Lambda_b(1/2^-)$ and $\Lambda_b(3/2^-)$ states. We find three $\Xi_b(1/2^-)$
and one $\Xi_b(3/2^-)$ states coming from the most attractive SU(6) $\times$
HQSS representations. Two of these predicted states, $\Xi_b(6035.4)$ and
$\Xi_b^*(6043.3)$, form a HQSS doublet similar to that formed by the
experimental $\Lambda_b(5912)$ and $\Lambda^*_b(5920)$ resonances. None of
these states have been detected yet, and their existence is also predicted by
constituent quark models. It constitutes a clear case for discovery.

\acknowledgments
We thank E. Oset and E. Ruiz Arriola for discussions and Anton Poluektov for
useful information on the LHCb experimental setup. This research was supported
by DGI and FEDER funds, under contracts FIS2011-28853-C02-02, FIS2011-24149,
FPA2010-16963 and the Spanish Consolider-Ingenio 2010 Programme CPAN
(CSD2007-00042), by Junta de Andaluc{\'\i}a grant FQM-225, by Generalitat
Valenciana under contract PROMETEO/2009/0090 and by the EU HadronPhysics2
project, grant agreement n. 227431. O.R. wishes to acknowledge support from
the Rosalind Franklin Fellowship. L.T. acknowledges support from Ramon y Cajal
Research Programme, and from FP7-PEOPLE-2011-CIG under contract
PCIG09-GA-2011-291679.

\end{document}